\documentclass[sigconf]{aamas} 
\setcopyright{none}
\usepackage{balance} % for balancing columns on the final page
\usepackage{amsfonts}
\usepackage{multirow}
\usepackage{multicol}
\usepackage{times}
\usepackage{enumitem}
\usepackage{latexsym}
\usepackage{graphicx}
\usepackage{amsmath}
\usepackage{booktabs}
\usepackage{multirow}
\usepackage{subfigure}
\usepackage{makecell}
\usepackage{algorithm}
\usepackage{algorithmicx}
\usepackage{hyperref}
\usepackage{listings}
\usepackage{csquotes}
\usepackage{bbding}
\usepackage{algpseudocode}
\usepackage[table,xcdraw]{xcolor}

\usepackage{enumitem}
\usepackage[T1]{fontenc}
\usepackage[utf8]{inputenc}

\usepackage{microtype}

\usepackage{graphicx}

\title[DebugTA: An LLM-Based Agent for Simplifying Debugging and Teaching]{DebugTA: An LLM-Based Agent for Simplifying Debugging and Teaching in Programming Education}

\author{
Lingyue Fu$^{\dagger,*}$,
Haowei Yuan$^{\dagger}$,
Datong Chen$^{\ddagger}$,
Xinyi Dai$^{\dagger}$, 
Qingyao Li$^{\dagger}$, \\
Weinan Zhang$^{\dagger}$,
Weiwen Liu$^{\dagger}$,
Yong Yu$^{\dagger}$ \Envelope
}

\affiliation{
$^{\dagger}$Shanghai Jiao Tong University, Shanghai, China\\
$^{\ddagger}$Fudan University
\city{Shanghai}
\country{China}
}
\email{* fulingyue@sjtu.edu.cn}

\begin{abstract}
 In programming education, Debugging and Teaching (DT) task is a common scenario where students receive assistance in correcting their erroneous code. The task involves multiple inputs, including erroneous code, error messages, reference solutions, and the question description, with the goal of generating modification suggestions to the erroneous code. However, two key challenges hinder the effectiveness of existing approaches. Firstly, the complexity and heterogeneity of inputs inherent in DT tasks significantly elevate the reasoning challenges faced by LLMs. Second, existing approaches often fail to fully leverage the availability of standard code in DT tasks, forcing models to rely solely on complex multi-step reasoning, which limits the potential of LLMs in addressing DT tasks effectively. To address these challenges, we propose DebugTA, a novel LLM-based debugging and teaching agent. DebugTA is equipped with a set of specialized tools, including a standard code retrieval tool, a variable substitution tool for generating an aligned reference code, and an external compiler interface for real-time code analysis and validation.  Guided by explicit pedagogical and debugging principles, DebugTA acts as an agent that decomposes a complex task into sequential LLM interactions, each utilizing distinct tools for specific subtasks, thereby simplifying the logical reasoning at each step and reducing overall reasoning complexity. Furthermore, DebugTA utilizes tool calls to align the standard code with the erroneous code as much as possible, allowing the LLM to focus on logic errors within the erroneous code and improving the accuracy of the generated suggestions. To rigorously assess the quality of modification suggestions, we introduce a student simulator-teacher interaction paradigm. Experimental results on three real-world code datasets demonstrate that DebugTA consistently improves teaching effectiveness while significantly reducing computational costs.
\end{abstract}

\newcommand{\BibTeX}{\rm B\kern-.05em{\sc i\kern-.025em b}\kern-.08em\TeX}

\begin{document}
\maketitle
\section{Introduction}

With the rapid advancement of large language models (LLMs), an increasing number of people are leveraging these models to support programming education~\cite{Yousef2025,Choi2024}. Programming-oriented LLMs, enhanced through retrieval-augmented generation (RAG)~\cite{du2024codegragbridginggapnatural} and fine-tuning techniques~\cite{guo2024deepseekcoderlargelanguagemodel}, have demonstrated remarkable performance in foundational tasks such as code generation and code completion. In the context of programming education, the introduction of educational elements brings about a new set of complex tasks~\cite{zhang2024simulatingclassroomeducationllmempowered, askarbekuly2024llm,al2024analysis} that existing programming-oriented LLMs are not equipped to handle effectively.

% \begin{figure}[t]
%     \centering
%     \includegraphics[width=0.9\linewidth]{figures/intro.pdf}
%     \caption{Demonstration of the Debugging and Teaching (DT) task. DebugTA processes erroneous code provided by students, compiler error messages,  standard code, and question descriptions to generate modification suggestions, thereby assisting students improve their code.}
%     \label{fig:intro}
% \end{figure}

\begin{figure*}
    \centering
    \includegraphics[width=0.8\linewidth]{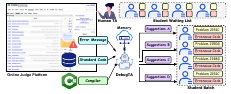}
    \caption{Demonstration of the Debugging and Teaching (DT) task. DebugTA receives the erroneous code and the standard code pool, connects to an external compiler, and then outputs tailored modification suggestions for each student.}
    \label{fig:intro}
\end{figure*}

Debugging and Teaching (DT) is a common scenario in which instructors guide students to identify and correct errors in their code. Formally, instructors provide guidance for modifications based on multiple inputs: erroneous code from the student, compiler error messages, the question description, and the standard code. During this interactive process, students engage with the instructors by adjusting their code according to the suggestions and resubmitting it for further evaluation. This process of learning from bugs and fixing them is a crucial learning path for students.

In practice, however, a single instructor often needs to support dozens of students, each with their own distinct coding style, which requires substantial effort to identify their bugs and provide effective guidance. This is especially common on Online Judge Platforms, such as LeetCode\footnote{https://leetcode.cn} and Codeforces\footnote{https://codeforces.com}, with numerous programming participants and insufficient human guidance. Recently, with the development of LLMs, LLMs have emerged as a potential method to assist in DT tasks, as illustrated in Figure \ref{fig:intro}. Even though some online judge platforms have deployed robot assistants to help with debugging, their adoption among students remains limited, primarily because the debugging suggestions provided by current LLM-based solutions are often inaccurate. This highlights the limitations of prompt-based approaches in effectively addressing the DT task. Nevertheless, LLM-based DT agents have the potential to substantially enhance the efficiency of computer science education by providing immediate, personalized feedback to students without requiring constant instructor intervention, if their accuracy and reliability can be improved.

% Challenge 1: high cost
One main challenge is that applying debugging methods to the DT task places substantial demands on the \textbf{reasoning capabilities of LLMs}.
Techniques such as LLM self-reflection~\cite{chen2023teachinglargelanguagemodels,LDB} and rethinking~\cite{li2024rethinkmctsrefiningerroneousthoughts} require models to perform advanced, multi-step logical reasoning to identify and correct errors. In addition, many of these methods rely on auxiliary resources such as compiler feedback~\cite{bi-etal-2024-iterative}, retrieve-augmented context~\cite{ALSAFARI2024100101}, or multi-agent collaboration~\cite{10807454}, which further increase the complexity of the reasoning process. Furthermore, the DT task involves highly heterogeneous inputs, including problem descriptions, erroneous code, and compiler error messages. When these inputs are presented simultaneously, LLMs often struggle to accurately process and interpret the relationships among them~\cite{zhang2023multilinguallargelanguagemodels,zhang2024codemixedllmimprovelarge}. This misalignment frequently increases the overall reasoning difficulty of the DT task and further leads to invalid or imprecise modification suggestions.

% Challenge 2: code mixing
% Another significant challenge in the DT task arises from the \textbf{code-mixing problem}, where the input data includes the description of the question, reference code, erroneous code, and error messages. This limitation stems from the syntactic and semantic differences between natural language and programming language, which requires LLMs to dynamically switch contexts during processing~\cite{zhang2024codemixedllmimprovelarge}. According to~\cite{zhang2023multilinguallargelanguagemodels}, the current multilingual capabilities of LLMs are not well-suited to handle code-mixing texts. Discrepancies between these input types can cause errors in understanding relationships and dependencies within the mixed input, leading to ineffective or incorrect suggestions in the DT task.

Another limitation is that current LLM-based debugging methods \textbf{neglect the usage of standard code} in the DT task. In the context of DT task, the presence of a standard code provides additional information that is not typically available in standard debugging tasks. This additional information can potentially simplify the reasoning process and enhance the quality of the generated suggestions. However, existing approaches that are directly adapted from debugging tasks fail to effectively utilize this standard code. It is important to note that leveraging reference code appropriately to enhance the accuracy of suggestions remains a non-trivial challenge. Simply concatenating standard and erroneous code as input may confuse the model and reduce the accuracy of suggestions~\cite{Dinh2023PotentialBug}. Additionally, it may lead to the leakage of the standard answer, potentially encouraging students to engage in plagiarism.

To address these challenges, we propose \textbf{DebugTA}, a \textit{\underline{Debug}ging and \underline{T}eaching LLM \underline{A}gent} that integrates debugging and teaching to help students correct their erroneous code. DebugTA is designed around a set of core tools, including a retrieval tool for selecting relevant reference code and an external compiler interface for real-time code validation. By autonomously retrieving appropriate reference solutions and aligning them with the student’s erroneous code, DebugTA enables subsequent LLM interactions to effectively leverage the correct algorithmic logic embedded in the standard code, rather than focusing merely on superficial stylistic or variable-level differences. Furthermore, by flexibly handling heterogeneous inputs through dedicated processing steps and decomposing complex reasoning into a sequence of targeted tool-assisted interactions, DebugTA substantially simplifies the logical reasoning required at each stage. This design not only mitigates the reasoning challenges inherent in DT tasks but also improves the overall accuracy and reliability of debugging suggestions.

To evaluate the effectiveness of DebugTA, we employ a student simulator to modify erroneous code based on the provided suggestions and regard the final accuracy of the modified code as the performance of these suggestions.
We evaluate DebugTA on three real-world code datasets of varying difficulty, with three different LLMs selected as backbone models for DebugTA.
Notably, our results indicate that DebugTA narrows the gap in teaching effectiveness between teacher models, as the agent’s stepwise approach alleviates the need for strong reasoning capabilities in each model.
The primary contributions of our work can be summarized as follows:
\begin{itemize}[leftmargin=10pt]
\item We introduce DebugTA, the first agent framework specifically tailored for the Debugging and Teaching (DT) task in programming education. DebugTA leverages correct reference solutions, a unique advantage of educational environments, and employs agent-driven decomposition along with a novel variable alignment mechanism to effectively utilize standard code and substantially reduce the reasoning burden on backbone LLMs.

\item We propose a student simulator–teacher agent interaction paradigm for a realistic and scalable evaluation of DT agents. Human validation experiments confirm that this paradigm faithfully simulates novice programmer behavior, providing a robust foundation for assessing automated debugging assistance.

\item We validate the effectiveness of DebugTA as a DT agent through comprehensive experiments on three real-world code datasets. Our implementation is publicly available at \url{https://anonymous.4open.science/r/DebugTA}.

\end{itemize}

\section{Related Work}

\subsection{LLMs in Programming Education}
LLMs have found broad applications in education~\cite{wang2024large, li2024adaptinglargelanguagemodels}. Research~\cite{10.1145/3613904.3642773, 10.1145/3626252.3630958} demonstrates the potential of LLMs in programming education. LLM-based agents have shown advantages in tasks such as program repair~\cite{10.1145/3650212.3680328} and example generation~\cite{10.1145/3636243.3636252}. Additionally, optimizations~\cite{10.1145/3634814.3634816} and innovations~\cite{TeachAIHowtoCode} in LLM-student dialogue models have proposed new solutions for programming education.
Meanwhile, \cite{app14104115} suggests that directly providing answers to students through LLMs may reduce their learning efficiency. This highlights the need to consider not only accuracy but also teaching efficiency when designing educational systems.

To study learning effectiveness, education researchers have conducted both simulated and real-world experiments. Real-world experiments are costly, and the number of student participants is often limited~\cite{10.1145/3657604.3662036}. Simulated students have been used in traditional AI education~\cite{ECSAL, li2023graph}. LLM-based student simulators further enhance this approach by incorporating characteristics~\cite{10.1145/3657604.3662031}, external memory~\cite{yue2025mathvcllmsimulatedmulticharactervirtual}, and training on historical records~\cite{xu2024eduagentgenerativestudentagents} to better model diverse student behaviors. Through simulated students, researchers can build virtual classrooms~\cite{zhang2024simulatingclassroomeducationllmempowered} to analyze the impact of instructional system designs on student learning. In this paper, we introduce a simulated student, \textit{StuBot}, for the DT task to evaluate teaching quality.

\subsection{Automated Debugging with LLMs}
Recent advancements have explored the application of LLMs in automated debugging. Methods such as self-reflection and static analysis have been used for bug identification and correction~\cite{LDB, kang2025explainable, yan2024betterdebuggingcombiningstatic}. By decomposing the debugging process, multi-agent systems have been designed to achieve high accuracy in debugging~\cite{yang2025coastenhancingcodedebugging, lee2024unifieddebuggingapproachllmbased}. Fine-tuning LLMs has also been employed to improve their debugging capabilities~\cite{jiang2024trainingllmsbetterselfdebug}. Despite the success of these approaches in debugging tasks, the complex input and reasoning of DT tasks remain challenging for these methods.

\section{Problem Definition}
% In this section, we first formally define the Debugging and Teaching (DT) task, followed by a description of the student simulator-teacher interaction paradigm, which is used to evaluate the effectiveness of the guidance provided by DebugTA.

\begin{figure*}[t]
    \centering
    \includegraphics[width=\textwidth]{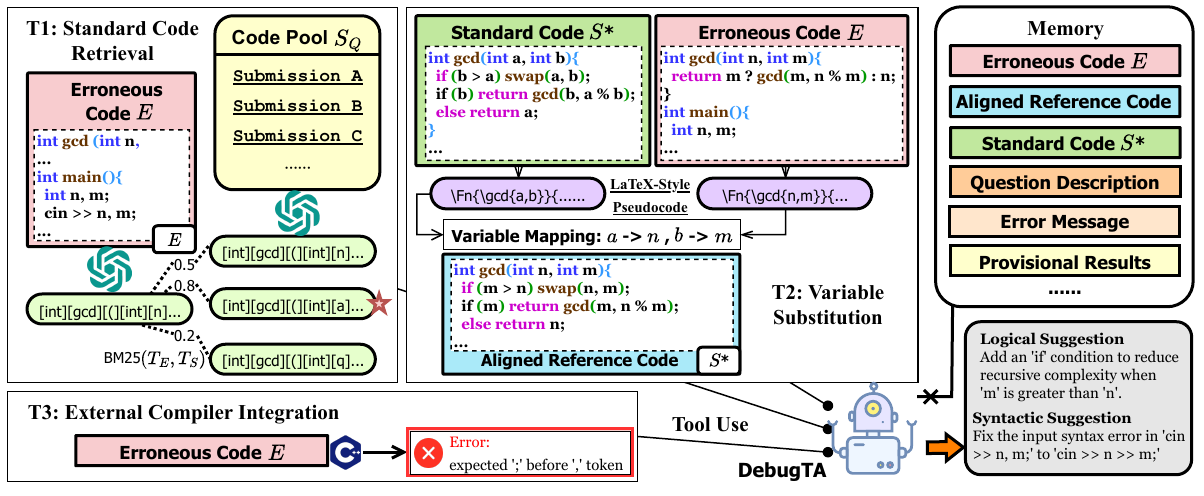}
    \caption{\textbf{Overview of DebugTA}. DebugTA is equipped with three specialized tools: Standard Code Retrieval, Variable Substitution, and Compiler, along with a memory module. DebugTA adaptively invokes these tools and manages information in memory to process erroneous code submissions, ultimately generating modification suggestions for students.
    }
    \label{fig:overview}
\end{figure*}
% TODO: change structural transformation-> syntactical
\subsection{Debugging and Teaching (DT)}
In the context of online judge platforms for programming education, a question is represented as a pair $ (Q, \mathcal{S}_Q) $, where 
$Q $ is the description of the question, and $\mathcal{S}_Q $ is the standard code pool, derived from correct solutions previously submitted by other students. The DT task involves execution results from the online judger, including error messages $ EM $ and scores, which are available to assist in the debugging process. However, testcases are not publicly available to students and instructors. Formally, given an erroneous program $ E $ and the corresponding problem $ (Q,\mathcal{S}_Q) $, the DT task aims to provide guidance to correct $ E $ by generating a set of modification suggestions $ M_E $. These suggestions help students identify and rectify errors in $ E $, improving both program correctness and their understanding of the question requirements.

\subsection{Student Simulator-Teacher Interaction}
Student simulators represent a widely adopted approach in AI-assisted education, serving as behavior models that emulate student responses and function as evaluation mechanisms for teaching tools~\cite{10.1145/3637528.3671872,ECSAL}.
As a widely established method~\cite{TeachAIHowtoCode,zhan2025coderagentsimulatingstudentbehavior,wu-etal-2025-embracing}, we propose a student simulator-teacher interaction paradigm and introduce an LLM-based student simulator, \textit{StuBot}, which interacts with DebugTA to iteratively refine erroneous programs. This simulation allows us to systematically assess the quality of debugging suggestions and the overall effectiveness of DT in an automated setting. We present both the formalized updating algorithm of \textit{StuBot} and its validation experiments in Appendix~\ref{app:student-teacher}.

Compared to experiments with human participants, using LLMs to simulate students offers two significant advantages: (1) The knowledge base and reasoning capabilities of LLMs (without internet access) remain consistent, allowing us to simulate a cohort of students with similar abilities. In contrast, human students exhibit varying levels of programming proficiency, which can lead to less convincing experimental results due to this heterogeneity. (2) \textit{StuBot} enables repeatable and cost-effective experiments on a scale. With each different response from the DT models, \textit{StuBot} can provide simulated student responses in real time, facilitating a comprehensive evaluation of numerous interaction scenarios that would be impractical with human participants.

% By simulating this iterative correction process, we can quantitatively evaluate how effectively DebugTA assists in debugging and teaching. Specifically, we measure the number of iterations required to reach a correct solution and analyze how different configurations of DebugTA impact StuBot’s progression. This simulated framework provides a controlled evaluation setting, allowing for systematic improvements in DT without direct reliance on human learners.

\section{Debugging and Teaching Agent}

Figure~\ref{fig:overview} presents the overall design of DebugTA. The agent is equipped with three tools specifically tailored for the DT task. During execution, DebugTA follows adaptive reasoning guidelines to strategically utilize these tools within each subtask, thereby minimizing the reasoning burden. Additionally, DebugTA incorporates an external memory to store various types of information. For each interaction with the LLM, only a relevant subset of this information is selected and provided as input, which reduces input complexity and alleviates reasoning difficulty. Finally, DebugTA generates comprehensive modification suggestions for students to revise their work and resubmit.

\subsection{Tool Use}\label{tooluse}
As an agent, DebugTA is equipped with three specialized tools that underpin its intelligent agent capabilities for the DT task.

\subsubsection{Standard Code Retrieval}  
This tool selects, from all standard code submissions on the OnlineJudge platform, the one whose implementation structure is most similar to the input code (typically the erroneous code requiring correction). DebugTA accesses a curated pool of reference solutions $\mathcal{S}_Q$, where each element represents a standard or accepted solution to a specific programming problem $Q$. On online programming practice platforms for algorithmic problem solving, $\mathcal{S}_Q$ is constructed from historical student submissions, official solutions, and community-contributed codes. For a given problem, $\mathcal{S}_Q$ typically contains a diverse set of solutions, reflecting multiple algorithmic paradigms (e.g., bubble sort, quicksort, or library-based approaches for sorting problems).

Given an erroneous code submission $E$, the retrieval tool uses token similarity based on BM25 to select the most relevant reference code $S^*$ from $\mathcal{S}_Q$:
$$
\text{CodeSearch}(E) \rightarrow S^*,
$$
where $S^*$ is the solution in $\mathcal{S}_Q$ with the highest similarity score to $E$. Detailed calculation of similarity score is presented in Appendix~\ref{app:bm25}. The similarity is measured primarily based on the structural and token-level correspondence between the submitted code and the reference solutions. This tool enables DebugTA to retrieve a reference solution that is structurally similar to the erroneous program $E$, thus facilitating context-aware and guideline-driven debugging support.

\subsubsection{Variable Substitution}

This tool is designed to align the variable naming between the reference standard code and the erroneous code. Variable Substitution takes the standard code $S^*$ and the erroneous code $E$ as input. The process consists of the following steps:
\begin{enumerate}[leftmargin=15pt]
    \item \textbf{Variable Mapping Generation:}
   The tool analyzes both $S^*$ and $E$ by converting them into Latex-style pseudocode representations. By comparing these representations, it automatically generates a variable mapping $M_V$, which establishes correspondences between variables in $S^*$ and those in $E$.
   \item \textbf{Variable Replacement:} Using $M_V$, the tool substitutes variable names in $S^*$ with their corresponding names from $E$, producing the aligned reference code $\tilde{S^*}$.

\end{enumerate}
   
The operation of the tool can be formalized as:
$$
\text{VarSubstitute}(S^*, E) \rightarrow \tilde{S^*},
$$
where $\tilde{S^*}$ is the version of $S^*$ with variable names replaced according to the mapping derived from $E$. The full prompts of this tool are shown in Appendix~\ref{app:prompts}.
This tool enables DebugTA to generate a reference code whose variable names are consistent with those in the erroneous program. Such alignment greatly simplifies subsequent logical comparison and correction steps, as it eliminates confusion caused by differing variable names and allows for more direct, context-aware debugging.

\subsubsection{External Compiler Integration}
The compiler tool provides real-time syntax validation by compiling the program $P$ and returning error messages $EM$. These messages support the identification and correction of syntax errors. The interface is:
$$
\text{Compile}(P) \rightarrow EM_P,
$$
where $EM_P$ contains the relevant compiler error information for $P$.
This tool allows DebugTA to verify intermediate code states and obtain additional diagnostic information, supporting more accurate and informative debugging decisions.

\subsection{Adaptive Reasoning Guideline}\label{reasoningsteps}
\paragraph{Memory}
DebugTA decomposes complex DT tasks into a series of simpler LLM interactions. Providing the entire context to the LLM does not reduce the complexity of the input; on the contrary, excessive information may introduce redundancy and further complicate the reasoning process. To address this, DebugTA is equipped with a memory module designed to read and write all relevant code and textual information. The memory is maintained in the form of a dictionary, allowing DebugTA to selectively retrieve specific code snippets or textual data at each step for LLM input. This selective context provision substantially reduces the reasoning difficulty for each interaction. This modular approach to context management ensures that each reasoning step remains focused and efficient, facilitating more accurate and effective suggestions.

Equipped with three specialized tools and a memory module, DebugTA follows an adaptive reasoning guideline that enables it to autonomously invoke these tools to accomplish the DT task. 

Upon receiving an erroneous program $E$, the agent first stores the erroneous program $E$ together with the problem description $Q$ in its memory. It then invokes the compiler tool to obtain the error messages $EM_E$.
When $EM_E \neq \emptyset$, this indicates the presence of syntactic errors in $E$. According to prior work~\cite{tian-etal-2024-debugbench}, existing LLMs are generally effective at resolving syntactic errors. In such cases, the agent provides both $E$ and $EM_E$ to the language model to generate syntactic correction suggestions and stores them in the memory:
$$
M_{ES} = \text{SynCorrection}(E, EM_E),
$$
where $M_{ES}$ denotes the set of syntax-level modifications.

When $EM = \emptyset$, the agent infers that $E$ contains logical errors rather than syntactic errors. In this scenario, DebugTA follows an adaptive reasoning guideline for logical correction, enabling it to dynamically select and invoke appropriate tools as needed. The process typically involves:
\begin{enumerate}[leftmargin=10pt]
    \item Invoking the \textbf{Standard Code Retrieval} tool to obtain a reference solution $S^*$ from the pool $\mathcal{S}_Q$, and storing $S^*$ in memory;
    \item Assessing whether variable alignment is necessary. If the agent determines that variable names differ significantly between $S^*$ and $E$, it invokes the \textbf{Variable Substitution} tool to generate the aligned reference code $\tilde{S^*}$ and stores it in memory. Otherwise, this step may be skipped;
    \item Retrieving $E$, the (potentially aligned) reference code, and the problem description $Q$ from memory, then generating logical correction suggestions, which are stored as $M_{EL}$:
    $$
    M_{EL} = \text{LogicCorrection}(\tilde{S^*}, E, Q).
    $$
\end{enumerate}
Throughout this process, DebugTA adaptively utilizes the compiler to verify the reliability of each intermediate result and implements remedial actions as needed. For example, if the agent detects that the standard code after variable substitution fails to pass the problem's test cases (an error during substitution), it will autonomously decide to re-initiate variable alignment or adjust the substitution strategy.

This adaptive, stepwise reasoning approach allows the agent to flexibly address the key challenges of the DT task. By dynamically selecting the most relevant tools and context at each step, the agent processes heterogeneous inputs in a structured yet flexible manner, effectively reducing input complexity and reasoning burden. This design leads to more accurate, interpretable, and efficient debugging assistance, fully aligning with the core objectives of DebugTA.

\section{Experiments}
Our experiments are designed to address the following research questions:

    \textbf{RQ1:} Does DebugTA effectively utilize the standard code and outperform other LLM-based methods on the DT task? 

   \textbf{RQ2:}  Does providing standard code to DebugTA lead to potential leakage of reference solutions?
   
    \textbf{RQ3:} Does DebugTA reduce the reliance on the underlying model's reasoning ability compared to baseline approaches?

    \textbf{RQ4:} What is the token cost incurred by DebugTA during the debugging process?

    \textbf{RQ5:} Is the student simulator–teacher interaction paradigm a reliable approach for evaluating model performance on the DT task? 
  
    \textbf{RQ6:} Are all tools in DebugTA's decomposition necessary and effective for the DT task?

    % \textbf{RQ6:} 

\subsection{Datasets and Setups}
We evaluate our proposed DebugTA using three datasets of varying difficulty, each containing real-world student submissions.
Detailed information about the datasets is summarized in Table~\ref{tab:dataset_info}.
The data collection methodology and preprocessing procedures are detailed in Appendix~\ref{app:datasetPrepro}.
% Detailed implementation details are demonstrated in Appendix~\ref{app:setups}.

We implement DebugTA using three different backbone models: GPT-4o-mini (API)~\cite{openai2024gpt4omini}, Qwen2.5-Coder-Instruct (7B)~\cite{qwen2.5}, and DeepSeek-Coder-V2-Lite-Instruct (16B)~\cite{deepseekcoderv2}, with GPT-4o-mini serving as the \textit{StuBot} (student simulator). We set the maximum number of interactions between \textit{StuBot} and DebugTA as $T_\text{max} = 3$. 
For reference code selection, to preprocess the erroneous program $E$ and the candidate solutions $S$, we employ GPT-2~\cite{gpt2} tokenizer for tokenization.
Our code is available in 
\href{https://anonymous.4open.science/r/DebugTA}{Anonymous Github} and will be publicly released after the paper is accepted.

\begin{table}[t]
    \caption{\textbf{Statistics of real-world datasets}. \textit{Pool Size}: average number of correct code submissions available for retrieval per question. \textit{Code Len.}: average number of lines in the erroneous code submissions.} 
    \label{tab:dataset_info}
    \centering
    \resizebox{\linewidth}{!}{ 
    \renewcommand\arraystretch{1.0}
    \setlength{\tabcolsep}{0.7mm}{
    \small
    \begin{tabular}{lcccc}
    \toprule
 \textbf{Dataset} &  \textbf{\# Questions} & \textbf{Pool Size} & \textbf{Code Len.} & \textbf{Difficulty} \\
 \midrule
 CodeApex & 109 & 11.84 & 17.1 & Easy \\
 ACMOJ & 100 & 85.12 & 83.7 & Hard \\
 Code4Bench & 87 & 78.71 & 58.3 & Hard\\
 \bottomrule
    \end{tabular}
    }
    }
\end{table}

% \subsection{Experimental}
% We implement DebugTA using three different backbone models: GPT-4o-mini (API)~\cite{openai2024gpt4omini}, Qwen2.5-Coder-Instruct (7B)~\cite{qwen2.5}, and DeepSeek-Coder-V2-Lite-Instruct (16B)~\cite{deepseekcoderv2}, with GPT-4o-mini serving as the StuBot (student simulator). We set the maximum number of interactions for StuBot and DebugTA as $T_\text{max} = 3$. 
% For reference code selection, to preprocess the erroneous program $E$ and the candidate solutions $S$, we employ GPT-2~\cite{gpt2} tokenizer for tokenization.
% Detailed implementation details are demonstrated in Appendix~\ref{app:setups}.
% Our code is available in 
% \href{https://anonymous.4open.science/r/DebugTA}{Anonmyous Github} and  will be publicly released after the paper is accepted.

\begin{table*}[t]
\centering

\renewcommand\arraystretch{1.0}
\setlength{\tabcolsep}{1mm}{
\small % 字体整体变小
\caption{Results of DebugTA and baselines on three datasets with three backbone LLMs. The best performance for each backbone model in each dataset is denoted in bold.}
\label{tab:mainresults}
\begin{tabular}{llccc|ccc|ccc}
\toprule
\multicolumn{2}{l}{} & \multicolumn{9}{c}{\textbf{Dataset}} \\
\cmidrule(r){3-11}
\multicolumn{1}{l}{\textbf{\makecell{Backbone \\ 
(\# Param.)}}} & \multicolumn{1}{l}{\textbf{ \ Teacher}} & \multicolumn{3}{c}{\textbf{CodeApex}} & \multicolumn{3}{c}{\textbf{ACMOJ}} & \multicolumn{3}{c}{\textbf{Code4Bench}} \\
\cmidrule(r){3-5} \cmidrule(r){6-8} \cmidrule(r){9-11}
 &  & \textbf{AC Rate} & \textbf{AC@all} & \textbf{Plag.} & \textbf{AC Rate} & \textbf{AC@all} & \textbf{Plag.} & \textbf{AC Rate} & \textbf{AC@all} & \textbf{Plag.} \\ 
\midrule
- & Origin $E$ & 15.72 & 0.00 & - & 27.02 & 0.00 & - & 12.41 & 0.00 & - \\
\midrule
& Direct Debug & 84.54 & 78.70 & - & 30.10 & 12.00 & - & 38.62 & 18.39 & - \\ 
& Debug with $S$ & 77.59 & 75.93 & 14.81 & 25.32 & 14.00 & 27.00 & 25.06 & 10.34 & 35.63 \\ 
& Self Debug (Expl.) & 73.29 & 71.30 & - & 35.15 & 16.00 & - & 34.02 & 10.34 & - \\ 
& Self Debug (Trac.) & 86.30 & 83.33 & - & 32.70 & 18.00 & - & 34.14 & 14.94 & - \\ 
& Direct Teach & 72.27 & 68.52 & 19.44 & 27.25 & 14.00 & 11.00 & 34.14 & 11.49 & 14.94 \\ 
\rowcolor{gray!30} 
\cellcolor{gray!0}
\multirow{-6}{*}{GPT-4o-mini} & {DebugTA} & \textbf{94.44} & \textbf{94.44} & \textbf{5.56} & \textbf{50.70} & \textbf{42.00} & \textbf{5.00} & \textbf{43.10} & \textbf{26.44} & \textbf{9.20} \\ 
\midrule
& Direct Debug & 69.21 & 65.74 & - & 24.50 & 7.00 & - & 27.70 & 10.34 & - \\ 
& Debug with $S$ & 63.61 & 61.11 & 15.74 & 21.95 & 4.00 & 12.00 & 22.30 & 6.90 & 29.89 \\ 
& Self Debug (Expl.) & 65.83 & 58.33 & - & 16.70 & 6.00 & - & 18.85 & 2.30 & - \\ 
& Self Debug (Trac.) & 72.64 & 65.74 & - & 21.15 & 3.00 & - & 24.25 & 3.45 & - \\ 
& Direct Teach & 73.80 & 70.37 & 11.11 & 29.10 & 12.00 & 8.00 & 35.06 & 14.94 & 10.34 \\ 
\rowcolor{gray!30} 
\cellcolor{gray!0}\multirow{-6}{*}{Qwen (7B)} & {DebugTA} & \textbf{87.96} & \textbf{86.11} & \textbf{7.41} & \textbf{42.55} & \textbf{25.00} & \textbf{3.00} & \textbf{38.39} & \textbf{18.39} & \textbf{3.45} \\ 
\midrule 
& Direct Debug & 75.97 & 70.37 & - & 30.40 & 13.00 & - & 23.68 & 6.90 & - \\ 
& Debug with $S$ & 47.96 & 46.30 & 46.30 & 6.70 & 4.00 & 66.00 & 15.52 & 10.34 & 75.86 \\ 
& Self Debug (Expl.) & 64.63 & 59.26 & - & 31.45 & 18.00 & - & 26.67 & 8.05 & - \\ 
& Self Debug (Trac.) & 70.79 & 65.74 & - & 27.65 & 13.00 & - & 27.47 & 9.20 & - \\ 
& Direct Teach & 62.96 & 61.11 & 32.41 & 21.30 & 11.00 & 37.00 & 28.39 & 12.64 & 25.29 \\ 
\rowcolor{gray!30} 
\cellcolor{gray!0}\multirow{-6}{*}{DeepSeek (16B)} & {DebugTA} & \textbf{91.30} & \textbf{90.74} & \textbf{4.63} & \textbf{42.10} & \textbf{24.00} & \textbf{9.00} & \textbf{39.20} & \textbf{20.69} & \textbf{4.60} \\ 
\bottomrule
\end{tabular}
}

\end{table*}

\subsection{Compared Baselines}
We evaluate the original erroneous program $E$ and use its performance as a basic baseline. In addition, we compare DebugTA against five baselines: 
\begin{enumerate}[leftmargin=15pt]
    \item \textbf{Direct Debug} uses basic prompts to perform a traditional debugging task, where the input consists of a description of the question $Q$ and the erroneous program $E$.
    \item \textbf{Debug with Standard Code} includes not only the question description $Q$ and erroneous program $E$, but also a reference solution $S^*$. 
    \item  \textbf{Self-Debug Explanation}~\cite{chen2023teachinglargelanguagemodels} implements rubber duck debugging \cite{Hunt2000}, where the model debugs the buggy code by explaining it line-by-line. The model analyzes the erroneous program $E$ by first explaining its implementation and then comparing it with the problem description $Q$. 
    \item  \textbf{Self-Debug Trace} \cite{chen2023teachinglargelanguagemodels} leverages the concept of execution tracing for debugging. In this approach, the model simulates the execution process of the erroneous program $E$ by generating step-by-step traces of how each line would execute, including intermediate variable states and control flow. 
    \item   \textbf{Direct Teach} follows a student simulator-teacher interaction paradigm, where the teacher receives the question description $Q$,
 the erroneous program $E$, and the reference solution $S^*$ in a single prompt. The \textit{StuBot} used is the same as DebugTA.

\end{enumerate}

Throughout the evaluation, we ensure that all baselines utilize the same LLM settings, testcases, and prompts formats with DebugTA. This
ensures a fair comparison and eliminates potential disruptions caused by changes in prompt formats.

\subsection{Metrics}
The effectiveness of the DT task is evaluated based on the quality of the code produced by StuBot after modification. Referring to~\cite{codeapex,DBLP:journals/corr/abs-2107-03374}, we use Accuracy Rate (AC Rate) and Rate of Passing All Test Cases (AC@all) as evaluation metrics. AC Rate represents the percentage of test cases passed, following the Olympiad in Informatics (OI) scoring mechanism. AC@all  follows a stricter evaluation criterion similar to the ACM-ICPC contest rules, where a solution is considered correct only if it passes all test cases. Together, AC Rate and AC@all provide both fine-grained and holistic perspectives on the performance of different models and methods on the DT task.

\paragraph{Plagiarism Analysis}
We also introduce \textbf{plagiarism rate} as a metric to detect potential plagiarism, preventing the LLM from directly releasing the standard code as a response. The plagiarism detection ensures the fairness of the evaluation. Additionally, in an education system, directly providing the standard code to students is highly detrimental, as it reduces students' engagement with the problem and significantly lowers their learning outcomes. To determine whether plagiarism has occurred, we calculate the similarity scores between the final tokenized code, the standard code, and the original erroneous code using the SequenceMatcher algorithm~\cite{ratcliff1988sequence}. The plagiarism check function assesses whether the final code is excessively similar to the standard code while ensuring sufficient divergence from the original erroneous code. If an answer is determined to be plagiarized, the score for that problem is set to 0. 
The detailed plagiarism detection algorithm is provided in Appendix~\ref{appendix:plagiarism}.

\subsection{Main Results (RQ1 and RQ2)}
We compare DebugTA with baseline debugging and teaching methods on three real-world datasets, with results in Table~\ref{tab:mainresults}. Our results indicate that DebugTA consistently performs best across all datasets and backbones. For more detailed insights, we present the experimental results and analysis categorized by error types in Appendix~\ref{app:errortype}.

More specifically, we summarize the following key observations: 

\textbf{(1) 
 Providing the standard code directly in the LLM prompt often degrades performance.} We observe that debugging with the standard code $S$ baseline always performs worse than the direct debug baseline. 
When both the erroneous code and unprocessed standard code are presented together, the LLM tends to focus on identifying superficial differences rather than performing a logical analysis of the problem statement. This finding highlights that simply introducing the standard code as a reference is insufficient; additional design considerations are necessary to effectively leverage standard code in the debugging process.

\textbf{(2) DebugTA's design for the DT task is effective.} Compared to direct teaching, DebugTA achieves an improvement ranging from 9.5\% to 45\%.  On the one hand, DebugTA decomposes the DT task into several independent tool invocations. Each tool call focuses on a specific subtask, thereby minimizing the reasoning difficulty required from the backbone LLM. On the other hand, DebugTA’s processing of standard code enables the LLM to extract useful information from the reference solution and generate more effective modification suggestions.

 \textbf{(3) The parameter size of the backbone LLM has minimal impact on DebugTA's effectiveness.} Across all experiments, DebugTA consistently outperforms all baselines, regardless of the backbone LLM. This advantage primarily arises from DebugTA's architecture, in which each tool invocation handles a relatively simple and well-defined subtask. As a result, even lightweight LLMs can complete these subtasks effectively, thereby reducing the operational costs of deploying DT systems on online education platforms. The parameter size dependence is further illustrated in Sec. \ref{sec:ModelSize}.

\textbf{(4) DebugTA effectively mitigates answer leakage compared to baselines that provide the standard code.} The two baselines that directly provide the standard code (Debug with $S$ and Direct Teach) exhibit significantly higher plagiarism rates than DebugTA, demonstrating that our designed pipeline substantially reduces the risk of answer leakage. This improvement is primarily due to our approach of retrieving and adaptively modifying the standard solution, aligning it more closely with the erroneous code. As a result, the model is better able to identify and correct specific errors, rather than simply copying the reference solution. In practical educational settings, reducing answer leakage ensures that students engage more deeply with the debugging process, which in turn promotes a better understanding of their mistakes and leads to improved learning outcomes.

 % This can be attributed to the modular design of the DebugTA, which decomposes the DT task into multiple subtasks. Each subtask is relatively simple, allowing even lightweight LLMs to effectively complete, significantly reducing the operational costs of online education platforms.

% \item \textit{DebugTA enables smaller models to refine the reasoning of larger models.} Notably, we use GPT-4o-mini as the backbone for StuBot, while DebugTA's smallest backbone model is Qwen-7B. Despite this, DebugTA successfully guides larger models by refining their reasoning through structured pipeline design. This insight suggests that through task decomposition and pipeline structuring, smaller models can effectively distill and enhance the performance of larger models, offering a novel approach to model efficiency and cost reduction in intelligent education systems.

\begin{figure}
    \centering
    \includegraphics[width=0.65\linewidth]{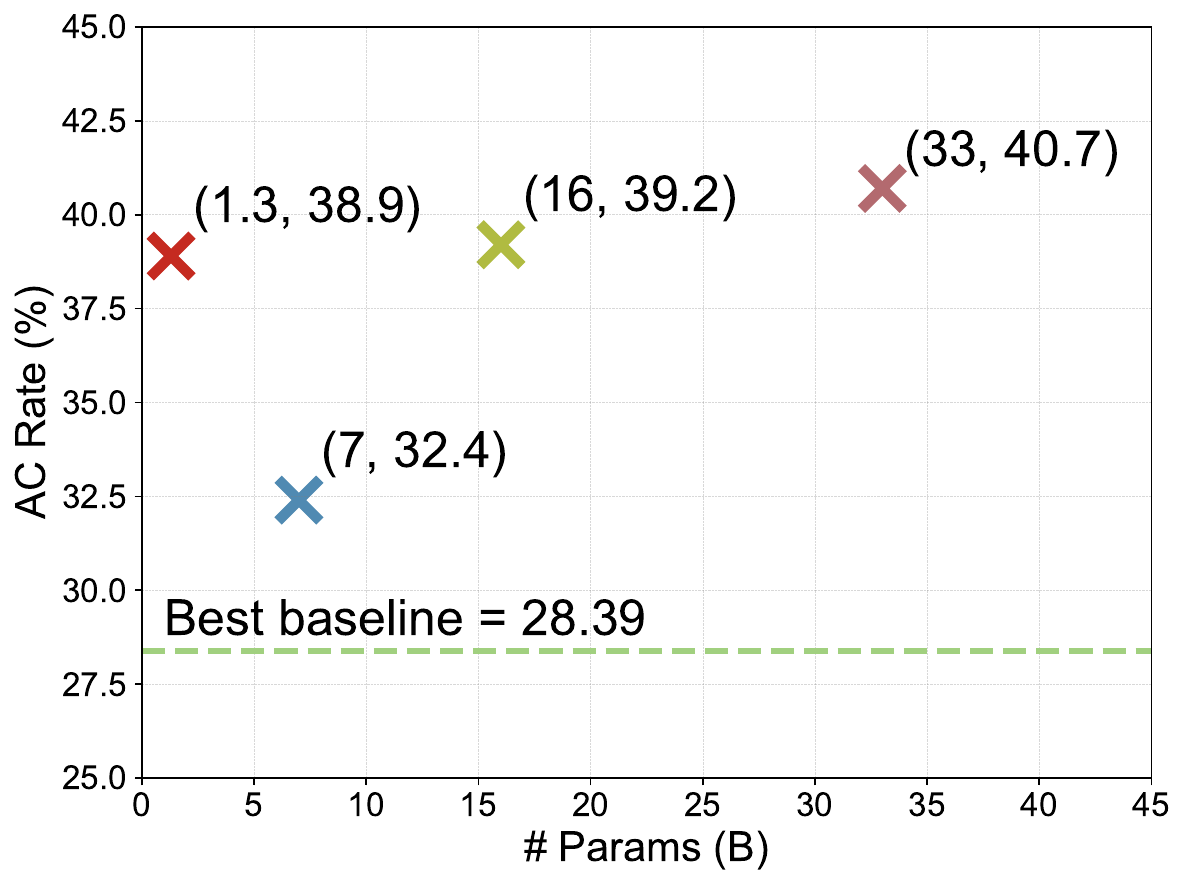}
    \caption{Performance of DebugTA using DeepSeek with various model sizes on the Code4Bench dataset. }
    % Backbones include DeepSeek-Coder-1.3B-Instruct, DeepSeek-Coder-7B-Instruct, DeepSeek-Coder-V2-Instruct (16B), and DeepSeek-Coder-33B-Instruct.}
\vspace{-10pt}
    \label{fig:modelsize}
\end{figure}

\begin{figure}
    \centering
    \includegraphics[width=0.7\linewidth]{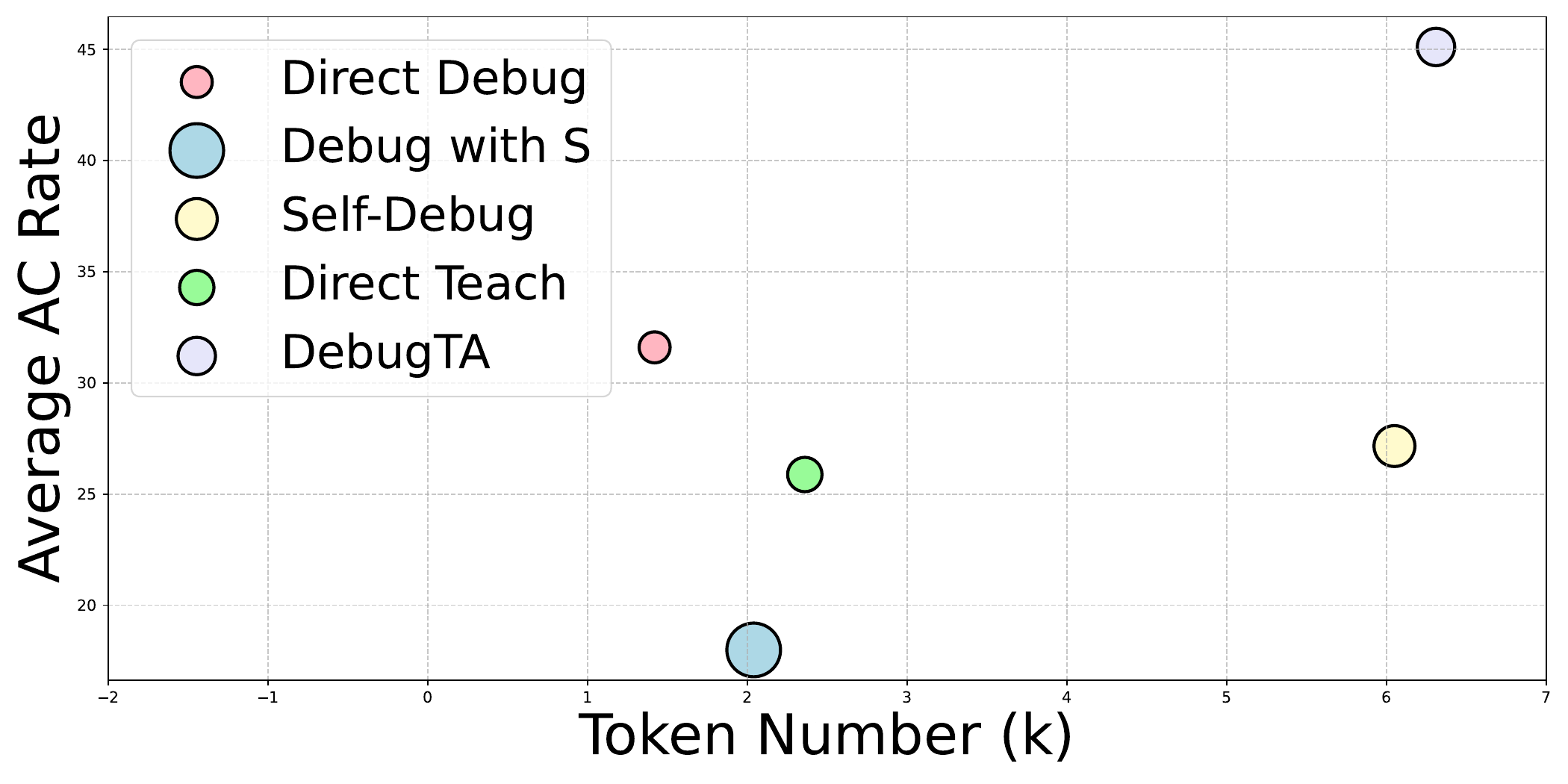}
    \caption{The relationship between token usage (x-axis), average AC rate on ACMOJ (y-axis), and performance variance (bubble size) between backbone LLMs.}
    \label{fig:tokens}
    \vspace{-10pt}
\end{figure}

\subsection{Performance on Different Model Size (RQ3)}\label{sec:ModelSize}
To demonstrate that DebugTA indeed reduces the reasoning difficulty required for the DT task, we investigate the impact of model size on the performance of the DebugTA framework and conduct experiments using DeepSeek with different numbers of parameters. Figure~\ref{fig:modelsize} illustrates the performance of DeepSeek-Coder-1.3B-Instruct, DeepSeek-Coder-7B-Instruct, DeepSeek-Coder-33B-Instruct and DeepSeek-Coder-V2-Instruct (16B) DebugTA on the Code4Bench dataset.

Our experimental results reveal that even small models outperform the best baseline, confirming that DebugTA effectively alleviates the dependence on large-scale reasoning capacity. Moreover, increasing the model size does not result in a notable performance improvement for the DebugTA framework. This is because DebugTA has minimal requirements for complex logical reasoning or advanced code generation abilities. Additionally, the experiment also indicates that improvements in an LLM's ability to generate code do not directly enhance its effectiveness as a programming teaching assistant.
\begin{figure}[t]
    \centering
    \includegraphics[width=\linewidth]{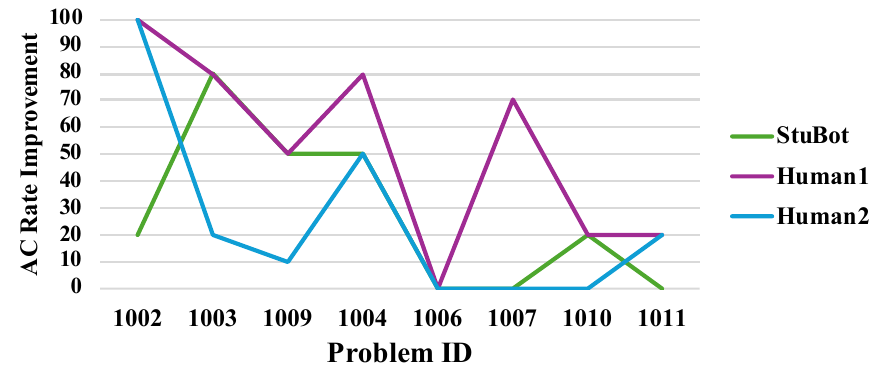}
    \caption{AC Rate improvement comparison between \textit{StuBot} and human participants across eight programming problems. Human 1 is a Ph.D. student in Computer Science and  Human 2 is a C++ beginner.}
    \label{fig:human}
    \vspace{-10pt}
\end{figure}

\begin{figure}[t]
    \centering
    \includegraphics[width=\linewidth]{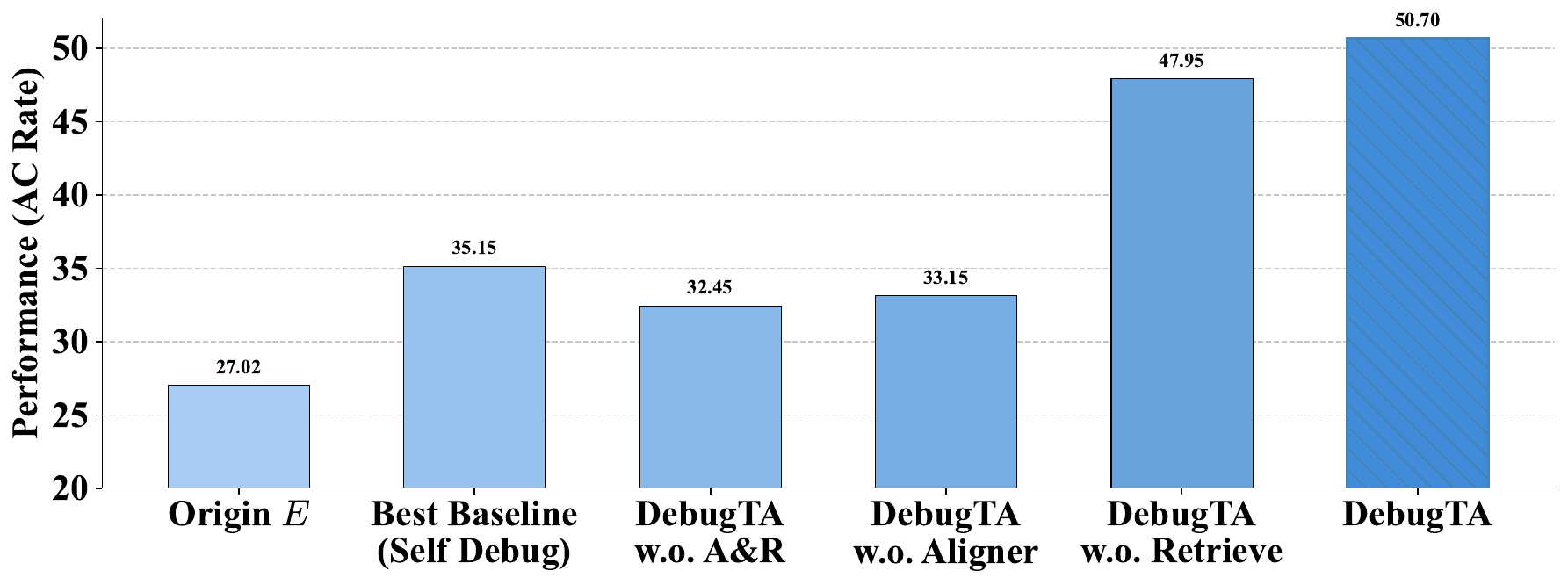}
    \caption{\textbf{Ablation Study Results on the ACMOJ}.}
    \label{fig:ablation_study}
    \vspace{-10pt}
\end{figure}

% \begin{table}[t]
%     \centering
%     \caption{\textbf{Token usage for baselines and DebugTA.}}
%     \label{tab:tokens}
%     \small % 字体整体变小
%     \begin{tabular}{lccc}
%     \toprule
%       \textbf{Avg. Tokens}   & \textbf{CodeApex} & \textbf{ACMOJ} & \textbf{Code4Bench}  \\\midrule
%        % Question  & 106 & 683 & 209\\
%        % Code $E$ & 111 & 721 & 442\\
%        % Code $\tilde{S^*}$ & 111 & 538 & 407\\
%        % Total &328 & 1.9k & 1.1k \\ \midrule
%        % Stage 1 & - & - & - \\
%        % Stage 2 & 1.40k & 2.83k & 2.44k \\
%        % Stage 3 & 1.27k & 3.48k & 2.20k\\ \midrule
%        Direct Debug & 0.24k & 1.42k & 0.67k \\
%        Debug with $S$ & 0.33k & 2.04k  & 1.16k \\
%        Self-Debug & 2.09k & 6.05k & 3.58k\\ 
%        Direct Teach & 0.65k & 2.36k & 1.48k \\
%        DebugTA & 2.66k & 6.31k & 4.64k\\
%         \bottomrule
% \end{tabular}
% \end{table}

\subsection{Token Usage Analysis (RQ4)}
To visualize the relationship between token consumption, performance, and stability across different methods, we present a bubble chart in Figure~\ref{fig:tokens}.
This visualization captures three key dimensions: token usage (x-axis), average performance score (y-axis), and performance variance (bubble size). 
Based on our calculations, DebugTA's token consumption consists of twice the input tokens plus a 2k prompt.  Although direct debugging/teaching methods use significantly fewer tokens, their performance is highly unstable and substantially inferior to DebugTA. Self-Debug approaches consume a similar number of tokens as DebugTA, but still underperform DebugTA. DebugTA's design effectively breaks down tasks and maximizes the use of standard code in the DT task, significantly reducing deployment costs in educational systems.
% As shown in Table~\ref{tab:tokens}, we present a comparison of token usage between baselines and DebugTA across different datasets. 
% Based on our calculations, DebugTA's token consumption consists of twice the input tokens plus a 2k prompt. Although direct debugging/teaching methods use significantly fewer tokens, their performance is highly unstable and substantially inferior to DebugTA. Self-Debug approaches (Explanation and Trace) consume a similar number of tokens as DebugTA. Compared to direct methods, they demonstrate more consistent performance across different backbone models and datasets, though still underperforming DebugTA. DebugTA's design effectively breaks down tasks and maximizes the use of standard code in the DT task, significantly reducing deployment costs in educational systems.

% we present the token usage statistics for the DT task input and the average tokens consumed at each stage by DebugTA.
% % Note that during the Reference Code Selection stage, DebugTA only utilizes GPT-2 for tokenization, and the generated tokens are not included in the token consumption.
% Based on our calculations, DebugTA's token consumption consists of twice the input tokens plus a 2k prompt. 
% Traditional debugging methods consume many tokens, such as LDB~\cite{LDB}, which uses 22-23k tokens per program. In contrast, DebugTA’s design effectively breaks down tasks and maximizes the use of standard code in the DT task, significantly reducing deployment costs in educational systems.

\begin{figure*}[t]
  \centering
\subfigure[CodeApex Dataset.]{\label{fig:round_codeapex}
\includegraphics[width=0.26\linewidth]{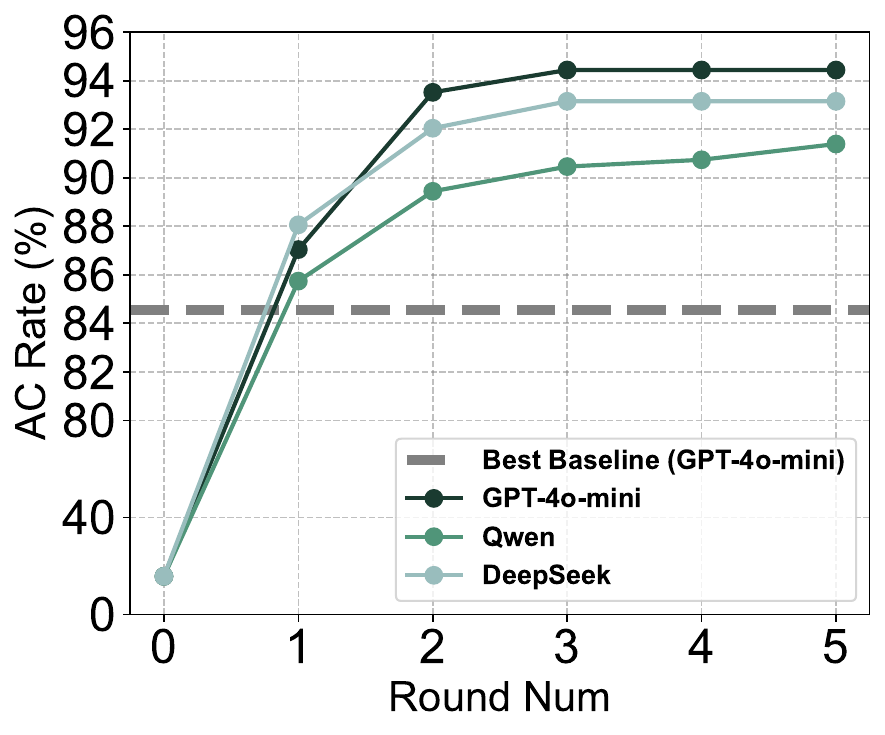}}
\subfigure[ACMOJ Dataset.]{\label{}\includegraphics[width=0.26\linewidth]{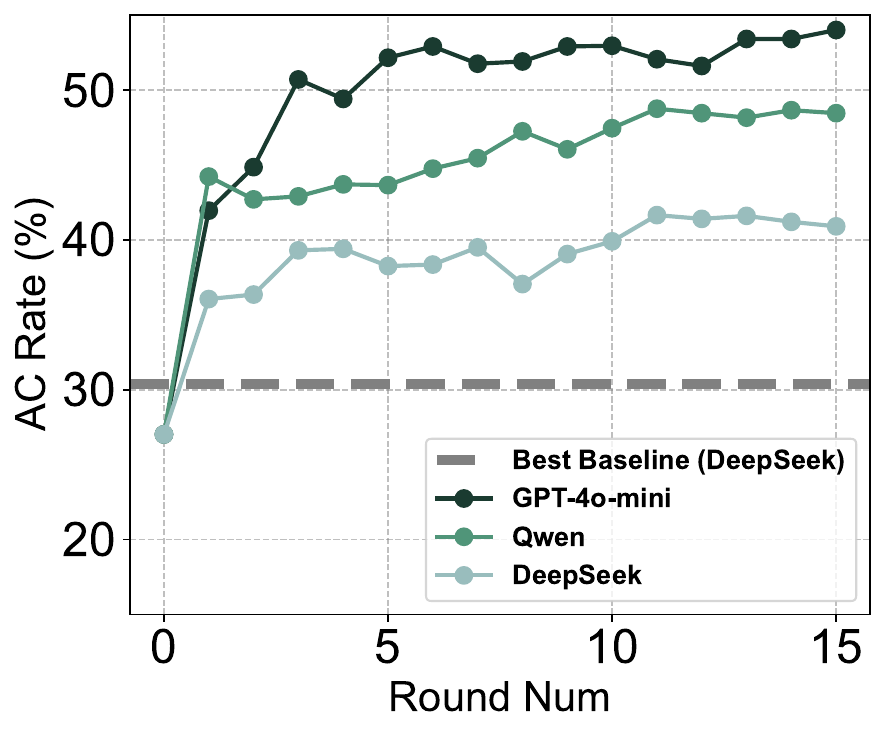}}
\subfigure[Code4Bench Dataset.]{\label{}\includegraphics[width=0.26\linewidth]{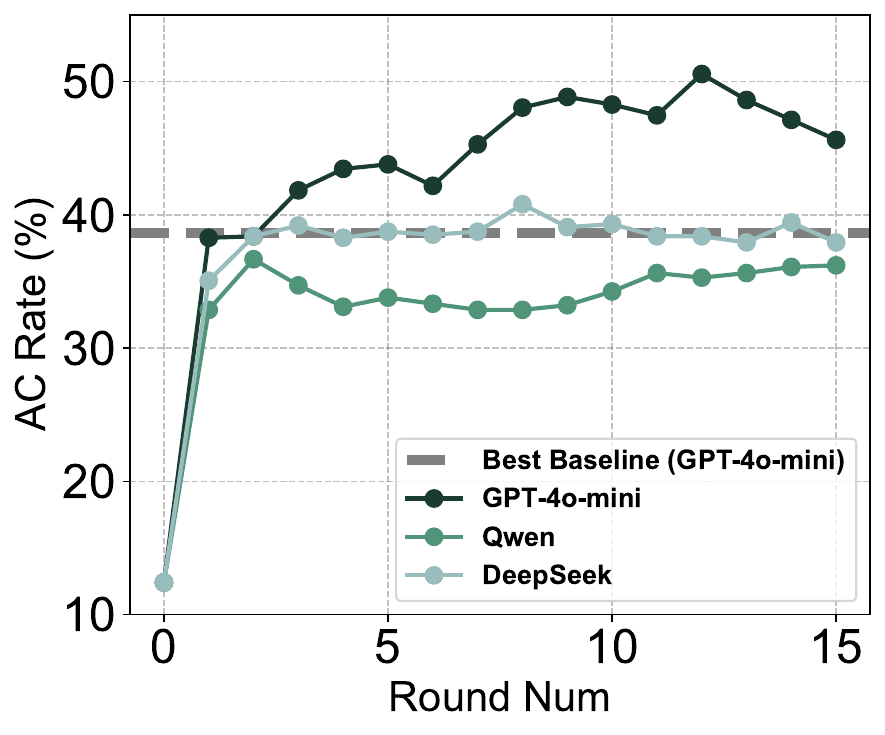}}
  \caption{Performance of DebugTA across three datasets with varying backbones, evaluated with progressively increasing interaction round numbers.}
  \label{fig:roundnum}
\end{figure*} 

\subsection{Human Participants Experiment (RQ5)}
To evaluate \textit{StuBot}'s ability to simulate human learners, we conduct an experiment with human participants. We recruited a Computer Science Ph.D. student (Human 1) and a C++ beginner (Human 2) to participate in a comparative study alongside \textit{StuBot}. All three subjects received identical problem descriptions $Q$, erroneous code $E$, and modification suggestions $M_E$, and are tasked with implementing appropriate code corrections.

The experiment consists of eight programming problems. The final AC Rate 
achieved by \textit{StuBot}, Human 1, and Human 2 are $46.25$, $71.25$, and $43.75$, respectively. Individual problem scores are illustrated in the Figure~\ref{fig:human}. Our analysis reveals that \textit{StuBot}'s performance trend closely aligns with that of human annotators. Human 1, possessing advanced programming knowledge, incorporated additional reasoning during the debugging process, consistently outperforming \textit{StuBot} across all problems. In contrast, Human 2 relied entirely on the modification suggestions provided by the LLM to implement corrections, resulting in performance slightly below that of \textit{StuBot}.

By employing \textit{StuBot} as a simulator for a student with a consistent knowledge level, we effectively mitigate the bias introduced by varying proficiency levels among different human learners. This approach enables more controlled and reproducible evaluation of debugging assistance strategies.

% \subsection{\fly{Human}}
\subsection{Ablation Study (RQ6)}\label{sec:Ablation}
To validate the effectiveness of each component in our proposed DebugTA framework, we conduct a comprehensive ablation study focusing on retrieving standard code $S^*$ and syntactical transformation through variable alignment. Figure~\ref{fig:ablation_study} illustrates the performance comparison across different variant configurations on the ACMOJ dataset. As shown in the figure, removing either the retrieval mechanism or the variable alignment module significantly impacts DebugTA's performance, with the alignment module having a particularly strong influence. These results provide an important insight: \textbf{providing LLMs with structurally and semantically similar correct code substantially reduces the complexity of the debugging task}, enabling more effective error identification.

\subsection{Impact of Round Number}
To investigate the relationship between DebugTA's performance and the number of interactions, we gradually increased the number of interactions between \textit{StuBot} and DebugTA. Figure~\ref{fig:roundnum} illustrates the performance of DebugTA across three datasets with varying backbones and interaction round numbers.

We observe that as the number of interactions increases, the accuracy of DebugTA improves. This indicates that DebugTA consistently offers valuable modification suggestions throughout the interaction process, helping students make necessary code corrections. After one round of interaction, DebugTA already outperforms the best baseline, highlighting the advanced design of the DebugTA pipeline. Furthermore, for simpler datasets, DebugTA requires fewer interaction rounds to guide students to optimal performance. In contrast, for more challenging datasets, the reasoning capabilities of the LLM remain critical, necessitating more interaction rounds to correct the code effectively.

% \begin{table}[h]
%     \centering
%     \begin{tabular}{lccc}
%        \toprule
%         & CodeApex & ACMOJ \\
%         \midrule
%         DebugTA & 94.44 & 50.70 \\
%         no Aligner & 93.52 & 47.95 \\
%         no Retrieve & 88.61 & 33.15 \\
%         no A\&R & 85.19 & 32.45 \\
%         \bottomrule
%     \end{tabular}
%     \caption{ablation}
%     \label{tab:ablation}
% \end{table}

% \subsection{Case Study}

\section{Conclusion}
In this paper, we introduce DebugTA, a novel LLM agent that integrates debugging and teaching for programming education. DebugTA utilizes three specialized tools along with a memory module to effectively address the challenges of high reasoning complexity and the underutilization of standard code, thereby streamlining the debugging process while preserving educational integrity. Through extensive evaluations on three real-world code datasets, we demonstrate that DebugTA enhances both the scalability and cost-efficiency of LLM-based programming education systems. Our results indicate that DebugTA not only improves the accuracy of the DT task, but also simplifies logical reasoning and reduces deployment costs for programming education.

\section*{Limitations}
One limitation of our approach arises when new problems are introduced into the teaching system and the standard code pool is relatively small. In such scenarios, the Standard Code Retrieval tool may struggle to find suitable reference solutions. A potential remedy is to employ a code generation LLM pipeline to automatically generate candidate solutions, then select those that pass all test cases to expand the standard code pool.

Meanwhile, in the teaching process, simply identifying the location of a bug for students is only the most basic step. The role of the instructor extends beyond this; teachers are expected to guide students in discovering the underlying causes of errors and in summarizing and generalizing problem-solving strategies. Currently, DebugTA is limited to pinpointing the bug location and does not yet possess the capability to facilitate deeper reflection or generalization of problem-solving approaches. In future work, we aim to develop a more advanced agent that can better emulate the comprehensive guidance provided by human instructors and to identify suitable experimental methodologies for evaluating the effectiveness of these enhancements.

% \begin{acks}
% \end{acks}

\newpage
\bibliographystyle{ACM-Reference-Format}
\bibliography{sample}

\clearpage
\appendix

\section{Details of {{StuBot}}}\label{app:student-teacher}

\subsection{Algorithm}
The interaction begins with an initial erroneous program $ E_0 $, which serves as the starting thought program for \textit{StuBot}. At each iteration, \textit{StuBot} receives a set of modification suggestions $ M_{E_t} $ from DebugTA based on its current program $ E_t $. \textit{StuBot} then integrates these suggestions to update its thought program, generating a revised version $ E_{t+1} $. DebugTA continues to analyze $ E_{t+1} $ and provides further refinement suggestions. This iterative process continues until either \textit{StuBot} generates a correct program or a predefined maximum number of interactions is reached. The process is formally described in Algorithm~\ref{alg:stuBot}.

\begin{algorithm}[h]
\caption{Student Simulator-Teacher Interaction Paradigm for DebugTA Evaluation}
\label{alg:stuBot}
\begin{algorithmic}[1]
\State \textbf{Input:} Erroneous program $ E $, Question Information $(Q, \mathcal{S}_Q)$, Maximum Iteration $T_\text{max}$
\State \textbf{Output:} Modified program
\State Initialize the thought of student simulator $ \textit{StuBot} $ with erroneous program $ E_0 = E$
\State Set $ t \gets 0 $
\While{program is incorrect \textbf{and} $ t \leq T_{\text{max}} $}
    \State Generate $M_{E_{t}}= $ DebugTA$(Q, \mathcal{S}_Q, E_t)$
    
    \State $E_{t+1} \gets$  \textit{StuBot} modifies $E_t$ based on  $ M_{E_t} $
    
    \State Set $E_{t+1}$  as thought of \textit{StuBot}
    \State Increment $ t \gets t + 1 $
\EndWhile
\end{algorithmic}
\end{algorithm}

\subsection{Prompts of \textit{StuBot}}
\begin{lstlisting}
You are a C++ student who has received an assignment. You've also been given error code and guidance notes. Now, you need to modify the error code according to the provided guidance.
Note: Please output the complete corrected code. The output format should be (```cpp ```).

Assignment requirements:
{QuestionText}
Error code:
```cpp
{Code}
```
Guidance:
{TeacherPrompt}
\end{lstlisting}

\section{BM25 Score in Standard Code Retrieval}\label{app:bm25}
In the Standard Code Retrieval, DebugTA uses the BM25 score~\cite{robertson2000experimentation} to evaluate the similarity between the token sequence $T_E$ of the erroneous code and each candidate solution $T_S$. The BM25 score is calculated as follows:
\begin{equation}
\begin{gathered}
\text{BM25}(T_E, T_S) = \sum_{t \in T_E} \text{IDF}(t) \frac{(k_1 + 1)f(t, T_S) }{f(t, T_S) + M},\\
M =  k_1\left[ (1 - b) + b \cdot \frac{|T_S|}{ \mathbb{E}_{S\in\mathcal{S}_Q} |T_S|}\right].
\end{gathered}\nonumber
\end{equation}
where $ f(t, T_S) $ is the term frequency of token $ t $ in token sequence $ T_S $, $ |T_S| $ is the length of token sequence $ T_S$, $ k_1 $ and $ b $ are hyper-parameters controlling the term frequency and length normalization. In this work, we set the hyper-parameter to $k_1 = 2.0, b=0.75$. Here $ \text{IDF}(t) $ is the inverse document frequency of token $ t $. It is usually calculated as
\begin{equation*}
\text{IDF}(t) = \log\left(\frac{|\mathcal{S}_Q| - n(t) + 0.5}{n(t) + 0.5} + 1\right),
\end{equation*}
where $|\mathcal{S}_Q|$ is the size of standard code pool $\mathcal{S}_Q$ and $n(t)$ is the number of the token sequence $T_S$ containing token $t$.

\section{Prompts of Variable Substitution}\label{app:prompts}
\begin{lstlisting}
# Prompt of converting programs into a LaTex-style pseudocode
Please generate pseudo code for the following C++ code.  
Format:{algorithm2e} package in latex. 
Note: Only output the pseudocode itself, do not output additional information (including \documentclass, \usepackage, \begin{document}, etc.).
Please write the algorithm name in the caption of the pseudocode.
Code:
```cpp
{code}
```
Algorithm name: {name}.
\end{lstlisting}

\begin{lstlisting}
# Prompts for variable alignment
You are an experienced C++ programming expert who has received two pseudocode versions of a task - one correct code and one incorrect code. Now, please identify the corresponding variable names between the correct and incorrect code. Variables in both codes have similar functions, but note the following points:
1. Avoid simply swapping the order of two variables
2. Focus only on variable names, avoid confusing different variable types, check variable types, and especially avoid naming conflicts after modifications
3. Don't change the meaning of variables, ensure the code remains correct, only match variable names.
4. Output only the variable correspondence between correct and incorrect code in JSON format, without any additional text or explanation, according to this structure:
{
  "correct code variable":"incorrect code variable"
}
After analyzing the purpose of code variables, please output only this correspondence relationship in JSON format.
Example:
Correct code:
    \begin{algorithm}
    \caption{algorithm name}
    \KwIn{$n$ is the range.}
    \KwOut{$sum$ is the sum of the range.}
    set sum = 0
    \For{$t = 1$ \KwTo $n$}{
        sum += t\;
    }
    Print $sum$;
    \end{algorithm}
Incorrect Code:
    \begin{algorithm}
    \caption{algorithm name}
    \KwIn{$M$ is the range.}
    \KwOut{$s$ is the sum of the range.}
    set s = 0
    \For{$i = 1$ \KwTo $M$}{
        s += i;
    }
    Print $s$;
    \end{algorithm}
Output:
    {
        "n":"M",
        "t":"i",
        "sum":"s"
    }
Now I give you your task.
Correct code:
{pseudocode of reference code}
Incorrect Code:
{pseudocode of erroneous code}
\end{lstlisting}

\begin{table*}[t]
    \centering
    \caption{Performance comparison of different backbone models on syntax and logic errors across three datasets. Numbers in parentheses indicate the count of examples in each category.}
        \label{tab:error type}
    \footnotesize
    \begin{tabular}{llcccccc}
        \toprule
        & & \multicolumn{2}{c}{CodeApex (108)} & \multicolumn{2}{c}{ACMOJ (100)} & \multicolumn{2}{c}{Code4Bench (87)} \\
        \cmidrule(lr){3-4} \cmidrule(lr){5-6} \cmidrule(lr){7-8}
        & Error Type & Syntax (0) & Logic (108) & Syntax (7) & Logic (93) & Syntax (18) & Logic (69) \\
        \midrule
        \multirow{2}{*}{Origin} & AC Rate & - & 15.72 & 0 & 29.91 & 0 & 15.65 \\
        & AC@all & - & 0 & 0 & 0 & 0 & 0 \\
        \midrule
        \multirow{2}{*}{GPT-4o-mini} & AC Rate & - & 96.30 & 15.71 & 53.66 & 46.67 & 40.58 \\
        & AC@all & - & 96.30 & 14.29 & 44.09 & 22.22 & 23.19 \\
        \midrule
        \multirow{2}{*}{Qwen (7B)} & AC Rate & - & 90.46 & 34.29 & 42.31 & 25.56 & 37.10 \\
        & AC@all & - & 88.89 & 8.57 & 23.66 & 11.11 & 15.94 \\
        \midrule
        \multirow{2}{*}{DeepSeek (16B)} & AC Rate & - & 93.15 & 15.71 & 41.08 & 39.44 & 39.13 \\
        & AC@all & - & 91.67 & 14.29 & 22.58 & 22.22 & 20.29 \\
        \bottomrule
    \end{tabular}

\end{table*}

\section{Dataset Preprocess}\label{app:datasetPrepro}
The CodeApex dataset is sourced from the CodeApex benchmark~\cite{codeapex}\footnote{https://apex.sjtu.edu.cn/codeapex}, which includes code correction tasks and contains real student erroneous programs. The problems in CodeApex are entry-level programming exercises.
The ACMOJ dataset consists of real programming student submissions and problems from an open-source online judge platform\footnote{https://acm.sjtu.edu.cn/OnlineJudge}, which provides a public judging API. The ACMOJ dataset includes a wide range of algorithmic questions.
The Code4Bench dataset~\cite{code4bench} is sourced from Codeforces\footnote{https://codeforces.com} and provides real submissions of students. Questions in Code4bench are sourced from competitive programming contests.

In all datasets, we ensure that the erroneous code \( E \) used for testing is the first submission made by the student in a series of submissions within a short time frame. This guarantees that the initial thought of \textit{StuBot}, \( E_0 = E \), reflects an early-stage solution that has not yet through subsequent corrections. Additionally, correct submissions made within the same frame as $E$ are excluded from the standard code pool to prevent the reference solution from being overly similar to the erroneous code.

\section{Plagiarism Detection Algorithm}
\label{appendix:plagiarism}

\textbf{Plagiarism Check} determines whether the final code overly resembles the standard code while maintaining sufficient divergence from the erroneous code. The algorithm is illustrated in Algorithm~\ref{alg:plagiarism}.
\begin{algorithm}[h]
\caption{Plagiarism Detection Algorithm}
\label{alg:plagiarism}
\begin{algorithmic}[1]
\Require Standard code $S^*$, Erroneous code $E$, Final code $F$  
\Require Similarity threshold $\tau_{\mathit{sim}} \in [0,1]$, Difference threshold $\tau_{\mathit{diff}} \in [0,1]$  
\Ensure Detection result $Plag \in \{\mathtt{True}, \mathtt{False}\}$  
\State Tokenize $S^*$, $E$, $F$ into sequences $T_S$, $T_E$, $T_F$  
\State Compute pairwise similarities:  
\[
\begin{aligned}
s_{\mathit{SF}} & \gets \mathrm{SequenceMatcher}(T_S, T_F) \\
s_{\mathit{EF}} & \gets \mathrm{SequenceMatcher}(T_E, T_F) \\
s_{\mathit{SE}} & \gets \mathrm{SequenceMatcher}(T_S, T_E)
\end{aligned}
\]  
\If{$s_{\mathit{SE}} > \tau_{\mathit{sim}}$}  
    \State $Plag \gets \mathtt{False}$ \Comment{Not Plagiarized}  
\ElsIf{$s_{\mathit{EF}} > \tau_{\mathit{sim}}$ \textbf{or} $s_{\mathit{EF}} > s_{\mathit{SF}}$}  
    \State $Plag \gets \mathtt{False}$ \Comment{Not Plagiarized}  
\ElsIf{$s_{\mathit{SF}} > \tau_{\mathit{sim}}$ \textbf{or} $s_{\mathit{SF}} > s_{\mathit{EF}} + \tau_{\mathit{diff}}$}  
    \State $Plag \gets \mathtt{True}$ \Comment{Plagiarized}  
\Else  
    \State $Plag \gets \mathtt{False}$ \Comment{Not Plagiarized}  
\EndIf  
\State \Return $D$  
\end{algorithmic}
\end{algorithm}

The algorithm checks for plagiarism by calculating the similarity between the Standard code $S^*$, {Erroneous code $E$}, and {Final code $F$}. Specifically, if the final code significantly resembles the standard code while showing low similarity with the erroneous code, the algorithm returns True, indicating plagiarism. If the final code is more similar to the erroneous code or shows insufficient similarity to the standard code, it returns False, indicating no plagiarism. Two hyper-parameters control the strictness of this process: the {Similarity threshold} limits the maximum allowable similarity to the standard code, while the {Difference threshold} defines the acceptable divergence between the erroneous and standard code.

The {Plagiarism Rate} is calculated as the proportion of tasks identified as {plagiarized} out of all evaluated tasks. A plagiarized code is assigned a score of {zero}.

\section{Analysis for Error Types}\label{app:errortype}
To provide a more nuanced understanding of DebugTA's performance across different debugging scenarios, we present the experimental results categorized by error types in Table~\ref{tab:error type}. Since our data is sourced from online judge systems where students typically complete local testing in their IDEs before submission, syntax errors are relatively uncommon compared to logical errors.
The results reveal several important patterns.
\begin{itemize}[leftmargin=10pt]
    \item DebugTA demonstrates balanced performance improvements across both syntax and logic errors. This balanced efficacy can be attributed to the Guided Correction Stage's differentiated modification patterns for distinct error types, which effectively reduces input complexity and mitigates performance losses typically associated with code-mixing problems. By treating different error types with specialized approaches, DebugTA achieves consistent improvements regardless of error category.
    \item An interesting observation is that different backbone LLMs show relatively similar performance in correcting logical errors, while exhibiting more pronounced variations in syntax error correction. This suggests that DebugTA's task decomposition effectively lowers the reasoning capability threshold required for complex debugging tasks, enabling even smaller parameter models to identify logical issues by leveraging the reference solutions. The architecture's strength lies in its ability to guide models of varying capabilities toward effective debugging outcomes, particularly for the more challenging logical errors.
\end{itemize}


%%%%%%%%%%%%%%%%%%%%%%%%%%%%%%%%%%%%%%%%%%%%%%%%%%%%%%%%%%%%%%%%%%%%%%%%

%%% The acknowledgments section is defined using the "acks" environment
%%% (rather than an unnumbered section). The use of this environment 
%%% ensures the proper identification of the section in the article 
%%% metadata as well as the consistent spelling of the heading.

% \begin{acks}
% If you wish to include any acknowledgments in your paper (e.g., to 
% people or funding agencies), please do so using the `\texttt{acks}' 
% environment. Note that the text of your acknowledgments will be omitted
% if you compile your document with the `\texttt{anonymous}' option.
% \end{acks}

%%%%%%%%%%%%%%%%%%%%%%%%%%%%%%%%%%%%%%%%%%%%%%%%%%%%%%%%%%%%%%%%%%%%%%%%

%%% The next two lines define, first, the bibliography style to be 
%%% applied, and, second, the bibliography file to be used.

% \bibliographystyle{ACM-Reference-Format} 
% \bibliography{sample}

%%%%%%%%%%%%%%%%%%%%%%%%%%%%%%%%%%%%%%%%%%%%%%%%%%%%%%%%%%%%%%%%%%%%%%%%

\end{document}